\documentclass[runningheads]{llncs}
\usepackage{graphicx}
\usepackage{listings} 
\begin{document}

%

\title{``ChatGPT Is Here to Help, Not to Replace Anybody'' - An Evaluation of Students’ Opinions On Integrating ChatGPT In CS Courses}

\titlerunning{``ChatGPT Is Here to Help, Not to Replace Anybody''}
%

\author{Bruno Pereira Cipriano\orcidID{0000-0002-2017-7511} \and
Pedro Alves\orcidID{0000-0003-4054-0792}}
\authorrunning{B. Pereira Cipriano et al.}
\institute{Lusófona University, COPELABS, Lisbon, Portugal 
\email{\{bcipriano,pedro.alves\}@ulusofona.pt}}


%
\maketitle              
\begin{abstract}
Large Language Models (LLMs) like GPT and Bard are capable of producing code based on textual descriptions, with remarkable efficacy. Such technology will have profound implications for computing education, raising concerns about cheating, excessive dependence, and a decline in computational thinking skills, among others.

There has been extensive research on how teachers should handle this challenge but it is also important to understand how students feel about this paradigm shift. In this research, 52 first-year CS students were surveyed in order to assess their views on technologies with code-generation capabilities, both from academic and professional perspectives.


Our findings indicate that while students generally favor the academic use of GPT, they don't over rely on it, only mildly asking for its help. Although most students benefit from GPT, some struggle to use it effectively, urging the need for specific GPT training. Opinions on GPT's impact on their professional lives vary, but there is a consensus on its importance in academic practice.

\end{abstract}

\section{Introduction}

Large Language Models (LLMs) have been shown to have the capacity to generate computer code from natural language specifications \cite{xu_systematic_2022,daun2023chatgpt,destefanis2023preliminary}. Currently, there are multiple available LLM-based tools which display that behaviour. Two examples of such tools are OpenAI's ChatGPT \footnote{https://chat.openai.com/} and Google's Bard \footnote{https://bard.google.com/chat}.



This has implications for Computer Science (CS) education, since students now have access to tools that can generate code to solve programming assignments~\cite{prather2023s}, with a degree of success which allows them to obtain full marks or close to it \cite{prather2023s,savelka2023thrilled,cipriano2023gpt,finnie2023my,Savelka_2023,reeves2023evaluating}. This raised discussion amongst the CS education community, due to the risk of producing low quality graduates, lacking fundamental skills such as problem solving and computational thinking. Moreover, while the education community has faced technological challenges before--- such as the scientific calculator, the Internet, Google and StackOverflow---, LLMs have the capacity to generate a significant amount of code, although possibly with some level of mistakes \cite{Savelka_2023,cipriano2023gpt}, making it an automation instead of a helper.




There has been extensive research into how computer science teachers should respond to LLMs, adapting their teaching methods, assessments, and more \cite{lau2023ban,denny_promptly_2023,daun2023chatgpt,leinonen2023comparing,liffiton2023codehelp,finnie2022robots,prather2023robots}. Some educators resist (or fight), contemplating ways to prevent students from using these tools, such as blocking access or employing detection tools for AI-generated text with questionable effectiveness~\cite{open-ai-chat-gpt-FAQ-detectors}. Others embrace this new paradigm, adapting exercises so that students are encouraged to make the most of LLMs, with presentations/discussions or non-text-based prompts~\cite{denny_promptly_2023}, or analysing the tool's capacity to help students~\cite{hellas_exploring_2023}. In any case, teachers have been assuming the (heavy) burden of adapting to this new reality.

However, it is also important to understand the students' perspectives on this topic. They may ponder whether they perceive any sense of threat, or just joy due to the perceived simplicity of passing programming courses effortlessly. They may even believe that the role of a teacher has been rendered obsolete with the constant availability of a virtual assistant.

In this study, we conduct a survey with 52 first-year CS students. The survey aims to understand what students think of LLMs with code generation capacity, both from an academic and professional perspective, and if they are able to take advantage of them without having received any formal training on the matter.




We investigate five research questions:

\textbf{RQ1}: What is the opinion of first-year students regarding the academic use of tools like ChatGPT?

\textbf{RQ2}: To what extent can first-year students take advantage of tools like ChatGPT without any specific or formal training?


\textbf{RQ3}: What is the impact of ChatGPT on the learning experience of first-year students?

\textbf{RQ4}: What is the impact of ChatGPT on current teaching practices?

\textbf{RQ5}: What is the opinion of first-year students regarding the influence of AI-based code generation tools (like ChatGPT) on their professional future?



This paper makes the following contributions:
\begin{itemize}
\item Presents the empirical results of a student survey (N=52) focused on their opinion about the academic and professional usage of tools like ChatGPT;
\item Presents recommendations based on the aforementioned survey's results.
\end{itemize} 

\section{Related work}


A few recent works have evaluated CS students' opinions on ChatGPT and similar tools~\cite{zastudil2023generative,prather2023robots,liffiton2023codehelp,singh2023exploring,rahman2023chatgpt,yilmaz2023augmented}. The authors of~\cite{zastudil2023generative} conducted semi-structured interviews with 18 participants (12 students and 6 instructors) in order to investigate the respective experiences and preferences regarding the use of LLMs in computing courses. Amongst other findings, they reported that 1) students were interested in having LLM-oriented classes (consistent with \cite{singh2023exploring} which surveyed 430 CS MSc students), 2) students and instructors were concerned about the trustworthiness of these tools, as well as the potential for students to become over-reliant on them, 3) instructors had less experience with LLMs than students, and, 4) students and instructors disagreed on adapting the assessment, with instructors wanting to give more weight to proctored evaluations and reduce the weight of lab exercises, while students considered that the weight of programming labs should not be changed, indicating that those materials are valuable to their understanding. In~\cite{prather2023robots}, researchers used surveys to evaluate the opinions of 57 instructors and 171 students. A significant amount (72\%, approximately 123) of those students were enrolled in computer science or other related majors (e.g. software engineering). Amongst other finding, this research found that 1) students had slightly more experience using generative AI for code generation than instructors, which is coherent with the findings of~\cite{zastudil2023generative}, 2) both instructors and students strongly agree that generative AI cannot replace human instructors, and 3) first-year students seem to prefer getting help from their peers, while upper-level students seem to prefer using generative AI over other resources. The authors of~\cite{yilmaz2023augmented} assessed 41 students' views on using ChatGPT for learning programming, noting benefits like quick responses and debugging aid, and drawbacks such as encouraging laziness and giving wrong answers. Finally, the authors of~\cite{liffiton2023codehelp} used a 3-question survey to assess the efficacy of CodeHelp, a tool that acts as an intermediate between the student and GPT, in order to prevent students from over-relying on the LLM. The survey yielded mostly positive student feedback regarding the usefulness of the tool. 

The study in~\cite{zastudil2023generative} has findings similar to ours, such as students' interest in having LLM-related contents in their courses. However, while that study was based on qualitative methods, our study follows a mostly quantitative approach and our sample size is also larger. Our study is also comparable with~\cite{prather2023robots}, although that study included a small percentage of students from degrees unrelated to CS, while our study is solely focused on CS students. Our study encompasses a broader range of questions than those in~\cite{yilmaz2023augmented}, which, while having a similar sample size, focuses on a thematic analysis of answers to two open-ended questions. Finally, although CodeHelp's study~\cite{liffiton2023codehelp} involved CS students, its focus was on evaluating CodeHelp itself rather than standard interactions with ChatGPT.







In summary, we are still on the early stages of researching CS students' opinions on LLM-oriented classes and related topics. Also, to the best of our knowledge, our survey is the first one to ask CS students about their perception on being assessed on LLM-related skills.

\section{Experimental Context}


This study was performed in the scope of a Data-Structures and Algorithms (DSA) course belonging to a Computer Engineering degree, during the 2022/23 school year. The course takes place in the second semester of the first year, which means that students have already been exposed to one semester of programming. The course is composed of theoretical and practical (or lab) classes and follows a mixed approach of exercise-based and project-based learning~\cite{lenfant2023project}, with students being required to work on both weekly assignments with small coding questions, as well as in a larger project which takes months to develop and tries to mimic real-world software development. At the end of the semester, students are asked to `defend' their project by performing some changes to its code, in order to demonstrate knowledge and mastery of it\footnote{Project defenses are in-person and proctored.}. Both the exercise and project components are supported by an Open-source Automatic Assessment Tool (AAT).


The course's project is typically a command line application that performs queries on a very large data set, provided in the form of multiple CSV files. The queries must be implemented using efficient data structures and algorithms. This year, we used data from the Million Song data set\footnote{http://millionsongdataset.com/}. The project is split in two parts: the first part is focused on reading and parsing the input files, while the second part is focused on implementing the queries.

In this course, the teachers allow and even encourage students to interact with ChatGPT in a responsible way. This is true both for the weekly exercises, as well as for the course's main project. However, a cautionary note was added to the beginning of the weekly assignments' texts advising students to attempt to solve the problems on their own before resorting to technological support (not only ChatGPT, but also Google, StackOverflow, and so on.).


Finally, this year's project explicitly asked students to use ChatGPT in one of the requirements, with the caveat that it would be forbidden to use ChatGPT during the project's defense.


\subsection{The ChatGPT requirement and exercise}

To encourage students to use a structured approach with ChatGPT, in an attempt to promote their critical thinking~\cite{naumova_mistake-find_2023}, they were asked to implement one of the project's requirements using ChatGPT. The requirement involved parsing one of the project's input files, which contained information associating songs with artists, as shown in Listing \ref{lst:partial_instructions}.




More concretely, they were instructed to approach the requirement in the following way: 1) prompt ChatGPT for help with the requirement, 2) prompt for an alternative solution, and, 3) analyse and compare the GPT-generated solutions and present a written description of their findings. They were also expected to integrate ChatGPT's code into their own project.
 




\begin{lstlisting}[caption={The ChatGPT requirement (partial instructions)}, label={lst:partial_instructions}, basicstyle=\scriptsize\ttfamily, breaklines=true,breakindent=0pt,frame=single,captionpos=b, float]
Each line of the song_artists file follows one of the following syntaxes depending on whether the song is associated with a single artist or multiple artists:
<Song ID> @ ['<Artist Name>']
<Song ID> @ "['<Artist Name>', '<Artist Name>', ...]"
Where:
<Song ID> is a String;
<Artist Name> is a String;
\end{lstlisting}


\section{Survey}






\subsection{Methodology}
\label{ref:Methodology}
To help answer the research questions, students were administered a structured anonymous questionnaire during a class of the DSA course, and towards the end of the semester. This questionnaire was composed of multiple questions of various types (e.g. 1-5, Yes/No, categorical, open-ended), with each question contributing to one RQ. The questionnaire concluded with an open-ended qualitative question asking students for their opinions on GPT. The answers to the open-ended questions were analysed by both authors of this research, who then agreed on the best-fitting RQ to place each of them.




Of a total of 154 enrolled and participating students, 52 (33.77\%) replied to the questionnaire. Participation was optional and no incentive was given. The survey responses are available online\footnote{https://zenodo.org/records/8433052}.

\subsection{Results}



This section presents our findings with regard to each Research Question (RQ).

\subsubsection{RQ1: What is the opinion/sentiment of first-year students regarding the academic use of tools like ChatGPT?}

To answer this research question, our survey had 4 quantitative questions and 1 qualitative question.



In the first question, we asked students \textbf{[Q1.1] What is your opinion on the authorization by teachers of the use of GPT in academic projects?} [Scale: 1-5]. Only 1.9\% (1) of students selected option 1 (`Strongly disagree'), and no student selected option 2. 13.5\% (7) selected the neutral position (i.e. 3), while 30.8\% (16) selected option 4. Lastly, option 5 ('Strongly agree') was selected by a majority of 53.8\% (28) of students.



\textbf{[Q1.2] In part 1 of the project, you were asked to interact with GPT. Did you?} [Yes/No]. 84.6\% (44) of students replied ``Yes'', while 15.4\% (8) students replied ``No''.


\textbf{[Q1.3] If you answered ``No'' to the previous question, what is the reason for not interacting with GPT?} [Open-ended]. The following question was only answered by the 8 students that replied ``No'' to the previous question. We identified two major topics within the students' answers: ``I didn't feel the need to'' (6 occurrences) and ``I didn't know how to'' (2 occurrences). 

\textbf{[Q1.4] If not for the exercise, would you still have used GPT?} [Yes/No]. This question tried to understand if the students would have used GPT even if there was no requirement for doing so. Most students, 69.2\% (36) indicated that they would have used GPT anyway, but 30.8\% (16) indicated that they would not have used GPT unless asked to do so. We found this somewhat surprising, since we expected a higher percentage of students to indicate that they would have used GPT anyway. However, some students may have interpreted the question as being related to using GPT to implement the requirement related with the previous question, and not generically, as we intended.


\textbf{[Q1.5] Do you think it makes sense to evaluate your ability to interact with GPT?} [Yes/No]. The goal of the fifth question was to understand if the students would be open to being evaluated with regard to their ability to use tools such as GPT. Students' opinions on this topic are somewhat divided, with 57.7\% (30) indicating that they think it does not make sense to evaluate their capacity to interact with GPT and 42.3\% (22) agreeing with the evaluation. The balance between ``Yes'' and ``No'' voters somewhat surprised us: we expected a higher prevalence of the ``No'' option, since most students usually have negative opinions about being evaluated.

\textbf{In conclusion...} The majority of students agree with the usage of GPT in academic contexts and a substantial fraction (32.3\%) even agree on being assessed on their usage. Student S21 says \textit{"ChatGPT is a faster and more organized search tool that helps students solve problems and doubts they have"}.
However, there exists a minority of students who are either against this practice or neutral about it. This may be related to concerns of abuse and misuse: \textit{"students (...) request total or partial resolution of the exercise by GPT, often without fully understanding how the generated code works"} (S44) and \textit{"many people use it without knowing even a little about the subject"} (S26).

If directed to do so, a vast majority of students (84.6\%) will use GPT in an assignment, and most students will use it even if it's not asked of them, but the percentage is lower (69.2\%). In general, it seems that students have positive opinions with regard to these tools and their academic usage. 

\subsubsection{RQ2: To what extent can first-year students take advantage of tools like ChatGPT without any specific or formal training?}


The goal of this RQ was to understand if students would `naturally' be able to take advantage of these tools without having any formal training. This RQ was composed of 3 quantitative questions.


The first question relevant for RQ2 was \textbf{[Q2.1] ``From 1 to 5, quantify how many interactions/prompts do you usually need to get results that you consider useful? Consider the average number per problem you've tried to solve with GPT.''} [Scale: 1-5]. No student selected option 1, `A single prompt'. Option 2, `A few prompts', was selected by 38.5\% (20) of students, making it the most selected option, closely followed by option 3, selected by 36.5\% (19) of students. Option 4, `Many prompts' was selected by 23.1\% (12) of students. Finally, option 5, `I usually can't get useful results' was selected by 1.9\% (a single student). This shows us that most students are usually able to get GPT to produce useful results, but a significant fraction needs many prompts, probably more than would be necessary. This could mean that the students are not being effective in their prompts and could benefit from prompting training. 

As for the second question, \textbf{[Q2.2] ``Has the average number of interactions/prompts been reduced over time?''}, where 1 means "No" and 5 means "Yes, substantially reduced", the option with more votes was number 3, which was selected by 40\% (21) of students. Options 4 and 5 followed, with 19.2\% (10) and 17.3\% (9), respectively. Finally, options 1 and 2 were both selected by 11.5\% (6) of students.

\begin{figure}[ht]
\centering
\includegraphics[width=0.8\linewidth]{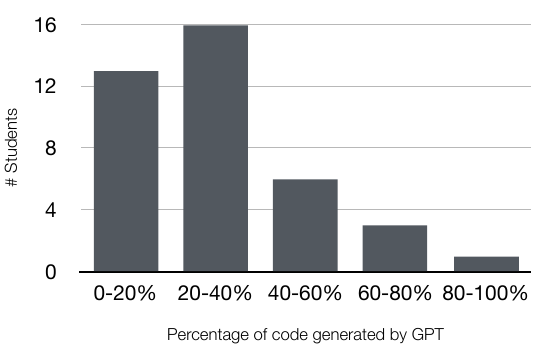}
\caption{Responses to the question ``[Q2.3] What percentage of the exercises' code was generated by
ChatGPT?''.}
\label{fig:chart-for-percentage-generated-gpt}
\end{figure}


\textbf{[Q2.3] ``In cases where you asked ChatGPT for help, what percentage of the exercises' code was generated by ChatGPT?''} Scale: Interval-based, ranging from [0-20\%[ to [80-100\%]. As depicted in Figure~\ref{fig:chart-for-percentage-generated-gpt}, the largest group of students, 41.06\% (16), fell within the 20-40\% range, closely followed by 33.33\% (13) who responded with values below 20\%. Still, a combined sum of 10.25\% selected options 60-80\% (3 students) and 80-100\% (1 student). These results suggest that most students seem to be using the tool to assist or supplement their coding process rather than relying on it completely. However, they also show that a few students tend to abuse the tool: using GPT to generate more than 40\% of an assignment probably means that the student is over-relying on it.


%

We decided to further investigate if the students that indicated the need for ``Many prompts'' (4) in Q2.1 also considered that they were improving over time. From the 12 students that selected that option, 1 student selected 1 (or ``No improvement''), 3 students selected 2, 6 students selected the middle option (3) and other 2 students selected option number 4. This shows that, even the students that have more difficulty (or report requiring more prompts), feel that they are improving somewhat, over time. However, only 2 out of those 12 students reported significant improvement.


\textbf{In conclusion...} Students' answers to these questions suggest that, in general, they are able to use GPT in a useful way. However, the average number of interactions needed seems to vary significantly between the students. This might be due to GPT not being deterministic, but it can also indicate that some students lack some `prompting skill`. Student S41 says: \textit{"Although ChatGPT is useful in some situations, it is not possible to obtain optimal results consistently"}.
Since some students (36.5\%) report improvements over time, we think that it is likely that `prompting skill` and/or `prompting experience' are at play here. That being said, if improvements are seen with unsupervised and untrained usage, then guided and trained usage could have the potential to yield even more improvements. As such, we believe that students would benefit from having some level of guidance provided by their teachers, in order to make them more efficient at interacting with these tools.

\subsubsection{RQ3: What is the impact of ChatGPT on the learning experience of first-year students?}

\begin{figure}[ht]
\centering
\includegraphics[width=1\linewidth]{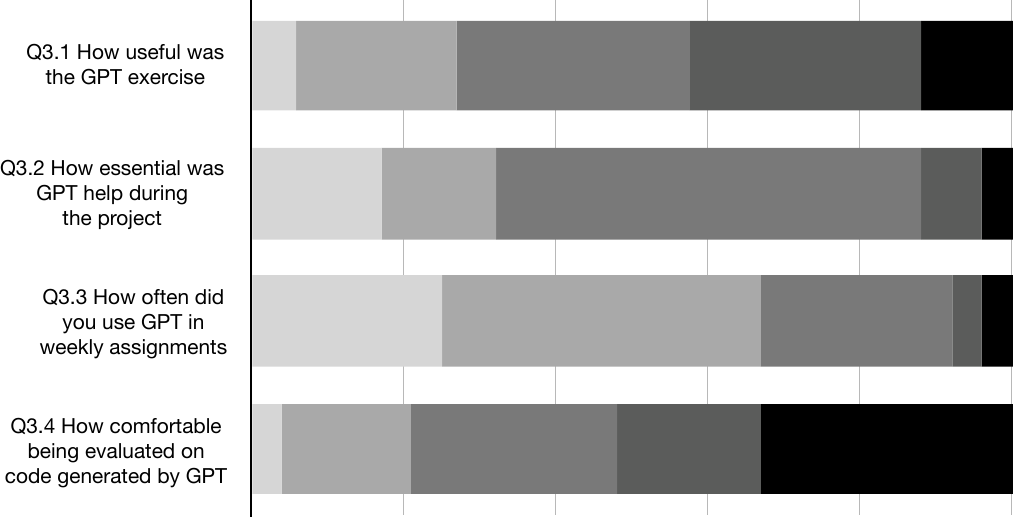}
\caption{Responses to some of the questions related to RQ3 (impact of ChatGPT on students' learning experience). The questions were on 1-5 scale, with 1 (lighter color) being the 'lesser' option and 5 (darker color) being the 'greater' option.}
\label{fig:chart-for-rq3}
\end{figure}


The goal of this RQ was to understand if tools such as ChatGPT have some kind of impact on the students' learning experience, \textbf{as per their own perception}. To answer this RQ, the questionnaire had 4 quantitative questions.

\textbf{[Q3.1] ``How useful do you think this exercise was (asking ChatGPT for help in processing the song\_artists file)?''} [Scale: 1-5]. As depicted in Figure~\ref{fig:chart-for-rq3}, option 1 (`Useless'), was selected by 5.8\% (3) of the students, while option 2 (`Slightly useful`) was selected by 21.2\% (11). Options 3 and 4 were both selected by the same percentage of students, with 30.8\% (16) of participants selecting each of them. Finally, option 5 (`Very useful'), was selected by 11.5\% (6) of students. These results show that most students considered the exercise useful, although opinions vary widely.


\textbf{[Q3.2] ``Regarding the project in general (parts 1 and 2). How essential has GPT's help been for you to do the project?''} [Scale: 1-5]. As illustrated in Figure~\ref{fig:chart-for-rq3}, 17.3\% (9) of students felt that GPT's assistance was `Not essential' (option 1), implying minimal to no reliance on the tool. A slightly smaller segment, 15.4\% (8), chose option 2, placing their reliance between `Not essential' and `Mildly essential'. A majority of respondent, 55.8\% (29) selected option 3 (or `Mildly essential'), suggesting they consulted GPT for assistance on some portions of their project. On the higher end of dependency, 7.7\% (4) leaned more towards frequent usage but not complete reliance. However, a very small fraction, 3.8\% (2), found GPT to be `Completely essential,' relying on it for almost the entirety of their project. In summary, while a majority of students used GPT occasionally for assistance, very few relied on it for the entire scope of their work.

\textbf{[Q3.3] ``How often did you use ChatGPT to help you with the weekly assignments?''} [Scale: 1-5]. As can be seen in Figure~\ref{fig:chart-for-rq3}, option 1, meaning `Never', was selected by 25\% (13) of students. The most selected option was option number 2, meaning `Only very sporadically', which was selected by 42.3\% (22) of students. Another 25\% (13) selected the mid-point (option 3). Option 4 was selected by 3.8\% (2) of students. Finally, at the extreme end of frequent adoption, another 3.8\% (2) of the students selected option 5, which was labeled `Always'. In summary, the majority of students either never used GPT in the weekly assignments or did so very sporadically. However, a few students seem prone to over reliance.




\textbf{[Q3.4] ``How comfortable are you with being evaluated (for example, in an oral or in a defense) about the code that was generated by GPT?''}. [Scale 1-5]. A minor 3.8\% (2) opted for the least comfortable option, which was labeled as `Not comfortable, I used the code and it works, but I don't understand what it does'. 17.3\% (9) students opted for the second option, while option 3, indicating a mid-range comfort level, was selected by 26.9\% (14). Option 4 was the choice for 19.2\% (10). Significantly, the most prominent response was option 5, labeled as 'Very comfortable, although I didn't write the code, I fully understand how it works', which was chosen by 32.7\% (17). This suggests a substantial portion of the participants felt highly comfortable in being evaluated in terms of their mastery of GPT's generated code, even if they hadn't written the code themselves. Refer to Figure~\ref{fig:chart-for-rq3} for a visualization of this question's reply distribution.

\textbf{In conclusion...}   
Students' opinions were divided about the usefulness of the specific GPT exercise. Even though most students tended to find it useful, there was a significant portion (27\%) that found the exercise with little or no usefulness. We hypothesize that these may be weaker students that couldn't get GPT to reach useful solutions, that they could use in their project.


A small fraction of students have shown a high dependency of GPT to implement the project (6 students or 11.5\%). On the other extreme, only 17,3\% indicated that they could have done the project without ever resorting to GPT for help, so the majority used it for occasional help.
Regarding weekly assignments, the reliance on GPT decreases, with only 4 students (7.6\%) admitting the need for full assistance. This led us to conclude that, although some students do abuse GPT, they are not the majority which only uses it sporadically. Interestingly enough, student S45 says: \textit{``ChatGPT is a useful tool but I think it is mostly useful for research and clarifying small doubts''}.



Finally, one of the most surprising results was related to assessment. Almost a third of the students feel very comfortable about being evaluated on code generated by ChatGPT. This suggests that students analyse and try to understand GPT-generated code rather than using it blindly.


\subsubsection{RQ4: What is the impact of ChatGPT on current teaching practices?}

This RQ tried to find the students' perceptions on how the current teaching methodologies and assessments should change because of GPT. Instead of understanding how teachers will change (the teacher's perspective), it is about understanding how students think the teachers should change (the student's perspective). Five questions were used to address this RQ.


\begin{figure}[ht]
\centering
\includegraphics[width=0.9\linewidth]{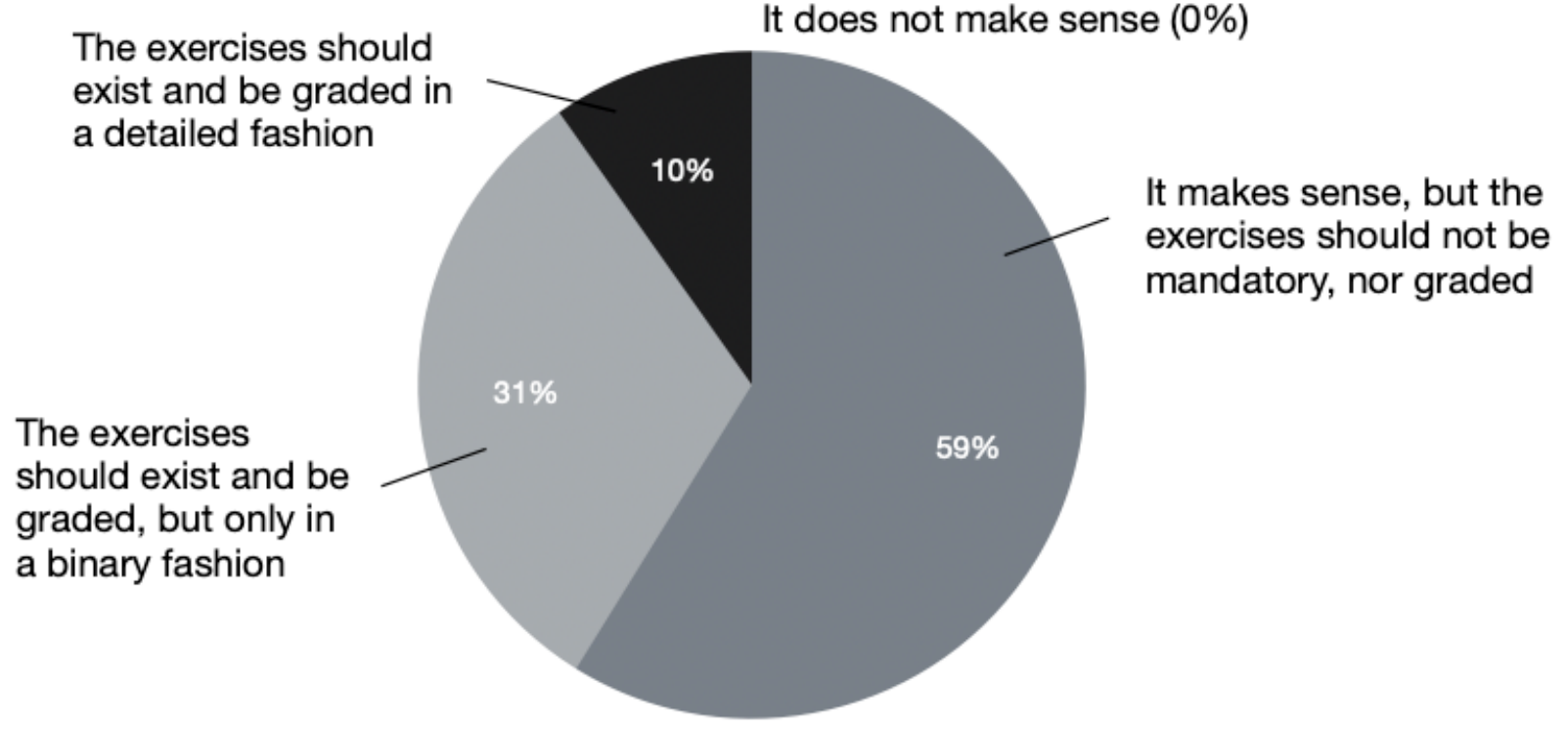}
\caption{Pie chart for question ``[Q4.1] Do you think it makes sense to include this type of exercises in the curriculum?''. All students agree with having GPT-based exercises, but they are divided in terms of evaluation scheme.}
\label{fig:chart-for-P17}
\end{figure}


\textbf{[Q4.1] Do you think it makes sense to include this type of exercises in the curriculum?} [Categorical]. This question aimed at understanding if students agree with having GPT-related contents in the course's curricula. As illustrated in Figure~\ref{fig:chart-for-P17}, all students think that it makes sense to include these types of exercises in the courses. 57.7\% (30) think the exercises should not be mandatory nor graded, 32.7\% (17) think that the exercises should exist and be graded, but only in a binary fashion (`Used GPT' / `Did not use GPT'), and finally, 9.6\% (5) think that the exercises should exist and be evaluated in a detailed fashion. All students agree on the inclusion of GPT-related exercises in university courses. However, there is a divergence in opinions regarding the assessment of interactions with GPT, as more than half of the students believe there should be no evaluation for such exercises.





Next, we asked students \textbf{[Q4.2] ``How helpful would it be for teachers to explain how to use GPT effectively (as they explain using the debugger, for example)?''} [Scale: 1-5]. No students selected the first option (`Not helpful'), 5.8\% (3) selected the second option, 17.3\% (9) selected the middle option, 25\% (13) selected the fourth option and, finally, a majority of 51.9\% (27) students selected the final option (`Very helpful'). These replies indicate that most students consider that having teachers explain how to use GPT would be useful.

The third question was a Yes/No question: \textbf{[Q4.3] ``Did the existence of GPT have any influence on your attendance at theoretical classes?''}. All students replied with `No'.


The fourth question was similar to the third one, but focused on the practical (or laboratory) classes: \textbf{[Q4.4] ``Did the existence of GPT have any influence on your attendance at practical classes?''}. The vast majority of the students (48 students, or 98.33\%) answered `No', while only 4 students (7.7\%) students answered `Yes'.

Finally, the last question was: \textbf{[Q4.5] ``Do you think ChatGPT could replace the teacher, as it ends up being a personal tutor?''}. Again, the vast majority of students answered `No' (50 students, or 96.2 \%), with only 2 students selecting `Yes'.

\textbf{In conclusion...} All students believe that these types of exercises should be included in the courses' curricula. However, there are significant divergences on whether these exercises should be graded, with most students (57.7\%) being against any form of grading.
Also, most students believe that it would be useful for teachers to explain how to use GPT. This is in line with the conclusions of RQ2: some students less efficient usage of GPT may be improved with formal training and teacher guidance. Students confirm this observation: \textit{``I think ChatGPT is a useful tool, but to do so it has to be well used and well understood. It would be great if there was help on how to use this tool from teachers''} (S26) and \textit{``[teachers should] explain to students more effective ways to use the tool and adapt to it''} (S39).

The availability of these tools is not having a significant influence in the students' class attendance, which is further confirmed by the fact that the vast majority of students (96.2\%) believe that GPT will not replace the teacher.  This result is consistent with the findings of~\cite{prather2023robots}.

\subsubsection{RQ5: What is the opinion of first-year students regarding the influence of AI-based code generation tools (like ChatGPT) on their professional future?}


This RQ was addressed through the use of 2 quantitative questions.


The first question was \textbf{[Q5.1] ``Will ChatGPT and other similar AI tools reduce the need for programmers?''} [Scale: 1-5]. Option 1, `Strong disagreement', was selected by 11.5\% (6) of students, while option 2 was selected by 21.2\% (11). The middle option was selected by 38.5\% (20) of students, making it the most frequently selected option. Options 4 and 5 were selected by 23.1\% (12) and 5.8\% (3) of students, respectively. As such, students are more or less divided on this topic.


Finally, students were asked \textbf{[Q5.2] ``Do you feel that the GPT-based exercises done in this curricular unit will make you more prepared for a possible future in which you have to interact professionally with GPT?''} [Scale: 1-5]. Only one student selected the first option ---`No, they will make it worse'--- while 9.6\% (5) students selected the second option---`Neutral: they will have no effect'. 30.8\% (16) of students selected the middle option, 34.6\% (18) selected the fourth option, and, finally, 23.1\% (12) selected the last option ---`Yes, very much'. Most students believe these exercises will have some importance to their careers if they have to interface with GPT professionally.


\textbf{In conclusion...} Students are not sure if GPT and similar AI code generating technologies will reduce the need for programmers, but they tend to agree that having GPT-based exercises in the DSA course will help them in their professional careers. One of the students said: \textit{``ChatGPT is here to help, not to replace anyone''} (S7), which inspired us for the title of the paper.

\section{Recommendations for CS Educators}

Students are likely to misuse tools like GPT-3.5 and Bard, which offer decent-to-good code generation for free. Also, LLMs are already being used by professional software developers \cite{barke2023grounded}, and this trend will possibly improve in the future as the models' capabilities advance further. This makes it important for students to get quality experiences with regard to interacting with these tools, so that they are aware of the tools' limitations and use them correctly. As such, we recommend that CS Educators start integrating these models in their courses. Following are some ideas derived from the survey results.

\textbf{Teach prompt-engineering techniques} Teach your students prompt engineering (PE) techniques, such as Role-playing~\cite{kong2023better} and Chain-of-Thought prompting~\cite{wei2022chain}.

\textbf{Employ LLM-based exercises} Design exercises in which students should interact with ChatGPT, Bard or similar tools. These exercises should be designed in order to promote skills such as critical thinking, code reading and understanding, and code critique. Exercises can range from prompt creation~\cite{denny_promptly_2023}, prompt improvement, assisted code generation and respective critique/comparison (such as our exercise), unit test generation, mistake finding~\cite{naumova_mistake-find_2023}, and so on. Also, they should be supervised by the teacher, who should guide and help students in this process, helping them identify problems with LLMs' output, thus promoting that students to not blindly trust the models.



\textbf{Evaluate students' prompting abilities} Consider evaluating students' prompting abilities and knowledge of PE techniques. Students appear open to be evaluated on this skill, which could be taken advantage of in order to promote their further development.

\section{Limitations}

The survey's authors are part of the DSA course teaching staff. This might have influenced some of the students' opinions.

As the survey participation was optional, there might have been some selection bias, and the students that decided to answer the questionnaire could be the ones that already had mostly positive opinions about these tools. Also, the students that participated were present in a theoretical class, which means that their opinions about attendance might not be representative of the general population.

GPT's behaviour is not deterministic and can also vary greatly over time~\cite{chen2023chatgpt}. This makes it hard to generalize some conclusions related with students' LLM interactions, since some students might have had better results than other students, not due to lack of `prompting skill', but due to the inherent variability in models themselves.



We expect most students to have used ChatGPT based on GPT-3.5, due to it being free. However, some students might have used the paid GPT-4 model, which was released at our location while our course was running\footnote{In Europe, GPT-4 became available for paying subscribers in March, 14, 2023}. Our results were not controlled for these model differences.

In relation to the questions about their present GPT-usage during weekly assignments and project (Q2.3, Q3.2 and Q3.3), it is conceivable that some students provided inaccurate responses in order to give us a false sense of security. However, we find this unlikely, since a significant number of students indicated they were comfortable in defending code generated by GPT.




\section{Conclusions}


Our study shows that students are generally in favor of using GPT in academic contexts, but they are divided regarding its assessment  (RQ1). Also, most students are already benefiting from GPT but some of them are not effective and efficient in its utilization, often requiring numerous prompts or failing to achieve a satisfactory solution (RQ2). Counter-intuitively, most students don't over-rely on GPT in the weekly assignments and project. Furthermore, a significant portion of students is comfortable with being evaluated on code generated by these tools (RQ3). The inclusion of specific GPT lessons and exercises is seen as beneficial and desirable (RQ4). Finally, students are considerably divided on the impact of GPT on their future professional careers, but in general, they agree that practicing its utilization during their academic journey is important (RQ5). 

\bibliographystyle{splncs04}
\bibliography{paper-survey}

\begin{thebibliography}{10}
\providecommand{\url}[1]{\texttt{#1}}
\providecommand{\urlprefix}{URL }
\providecommand{\doi}[1]{https://doi.org/#1}

\bibitem{barke2023grounded}
Barke, S., James, M.B., Polikarpova, N.: {Grounded Copilot: How Programmers Interact with Code-Generating Models}. Proceedings of the ACM on Programming Languages  \textbf{7}(OOPSLA1),  85--111 (2023)

\bibitem{chen2023chatgpt}
Chen, L., Zaharia, M., Zou, J.: {How is ChatGPT's behavior changing over time?} arXiv preprint arXiv:2307.09009  (2023)

\bibitem{cipriano2023gpt}
Cipriano, B.P., Alves, P.: {GPT-3 vs Object Oriented Programming Assignments: An Experience Report}. In: Proceedings of the 2023 Conference on Innovation and Technology in Computer Science Education V. 1. pp. 61--67 (2023)

\bibitem{daun2023chatgpt}
Daun, M., Brings, J.: {How ChatGPT Will Change Software Engineering Education}. In: Proceedings of the 2023 Conference on Innovation and Technology in Computer Science Education V. 1. pp. 110--116 (2023)

\bibitem{denny_promptly_2023}
Denny, P., Leinonen, J., Prather, J., Luxton-Reilly, A., Amarouche, T., Becker, B.A., Reeves, B.N.: Promptly: {Using} {Prompt} {Problems} to {Teach} {Learners} {How} to {Effectively} {Utilize} {AI} {Code} {Generators} (Jul 2023), \url{http://arxiv.org/abs/2307.16364}, arXiv:2307.16364 [cs]

\bibitem{destefanis2023preliminary}
Destefanis, G., Bartolucci, S., Ortu, M.: {A Preliminary Analysis on the Code Generation Capabilities of GPT-3.5 and Bard AI Models for Java Functions}. arXiv preprint arXiv:2305.09402  (2023)

\bibitem{finnie2022robots}
Finnie-Ansley, J., Denny, P., Becker, B.A., Luxton-Reilly, A., Prather, J.: The robots are coming: Exploring the implications of openai codex on introductory programming. In: Proceedings of the 24th Australasian Computing Education Conference. pp. 10--19 (2022)

\bibitem{finnie2023my}
Finnie-Ansley, J., Denny, P., Luxton-Reilly, A., Santos, E.A., Prather, J., Becker, B.A.: My ai wants to know if this will be on the exam: Testing openai’s codex on cs2 programming exercises. In: Proceedings of the 25th Australasian Computing Education Conference. pp. 97--104 (2023)

\bibitem{hellas_exploring_2023}
Hellas, A., Leinonen, J., Sarsa, S., Koutcheme, C., Kujanp{\"a}{\"a}, L., Sorva, J.: Exploring the {Responses} of {Large} {Language} {Models} to {Beginner} {Programmers}' {Help} {Requests} (Jun 2023). \doi{10.1145/3568813.3600139}

\bibitem{kong2023better}
Kong, A., Zhao, S., Chen, H., Li, Q., Qin, Y., Sun, R., Zhou, X.: Better zero-shot reasoning with role-play prompting. arXiv preprint arXiv:2308.07702  (2023)

\bibitem{lau2023ban}
Lau, S., Guo, P.: {From "Ban it till we understand it" to "Resistance is futile": How university programming instructors plan to adapt as more students use AI code generation and explanation tools such as ChatGPT and GitHub Copilot} (2023)

\bibitem{leinonen2023comparing}
Leinonen, J., Denny, P., MacNeil, S., Sarsa, S., Bernstein, S., Kim, J., Tran, A., Hellas, A.: Comparing code explanations created by students and large language models. arXiv preprint arXiv:2304.03938  (2023)

\bibitem{lenfant2023project}
Lenfant, R., Wanner, A., Hott, J.R., Pettit, R.: Project-based and assignment-based courses: A study of piazza engagement and gender in online courses. In: Proceedings of the 2023 Conference on Innovation and Technology in Computer Science Education V. 1. pp. 138--144 (2023)

\bibitem{liffiton2023codehelp}
Liffiton, M., Sheese, B., Savelka, J., Denny, P.: {CodeHelp: Using Large Language Models with Guardrails for Scalable Support in Programming Classes}  (2023). \doi{10.1145/3631802.3631830}

\bibitem{naumova_mistake-find_2023}
Naumova, E.N.: A mistake-find exercise: a teacher{\textquoteright}s tool to engage with information innovations, {ChatGPT}, and their analogs. Journal of Public Health Policy  \textbf{44}(2),  173--178 (Jun 2023). \doi{10.1057/s41271-023-00400-1}

\bibitem{open-ai-chat-gpt-FAQ-detectors}
OpenAI: How can educators respond to students presenting ai-generated content as their own? \url{https://help.openai.com/en/articles/8313351-how-can-educators-respond-to-students-presenting-ai-generated-content-as-their-own} (2023), [Online; last accessed 03-October-2023]

\bibitem{prather2023robots}
Prather, J., Denny, P., Leinonen, J., Becker, B.A., Albluwi, I., Craig, M., Keuning, H., Kiesler, N., Kohn, T., Luxton-Reilly, A., MacNeil, S., Peterson, A., Pettit, R., Reeves, B.N., Savelka, J.: {The Robots are Here: Navigating the Generative AI Revolution in Computing Education} (2023). \doi{10.1145/3623762.3633499}

\bibitem{prather2023s}
Prather, J., Reeves, B.N., Denny, P., Becker, B.A., Leinonen, J., Luxton-Reilly, A., Powell, G., Finnie-Ansley, J., Santos, E.A.: {“It’s Weird That it Knows What I Want”: Usability and Interactions with Copilot for Novice Programmers}. ACM Transactions on Computer-Human Interaction  \textbf{31}(1),  1--31 (2023)

\bibitem{rahman2023chatgpt}
Rahman, M.M., Watanobe, Y.: {ChatGPT for Education and Research: Opportunities, threats, and Strategies}. Applied Sciences  \textbf{13}(9), ~5783 (2023)

\bibitem{reeves2023evaluating}
Reeves, B., Sarsa, S., Prather, J., Denny, P., Becker, B.A., Hellas, A., Kimmel, B., Powell, G., Leinonen, J.: Evaluating the performance of code generation models for solving parsons problems with small prompt variations. In: Proceedings of the 2023 Conference on Innovation and Technology in Computer Science Education V. 1. pp. 299--305 (2023)

\bibitem{savelka2023thrilled}
Savelka, J., Agarwal, A., An, M., Bogart, C., Sakr, M.: {Thrilled by Your Progress! Large Language Models (GPT-4) No Longer Struggle to Pass Assessments in Higher Education Programming Courses}  (2023). \doi{10.1145/3568813.3600142}

\bibitem{Savelka_2023}
Savelka, J., Agarwal, A., Bogart, C., Song, Y., Sakr, M.: {Can Generative Pre-trained Transformers ({GPT}) Pass Assessments in Higher Education Programming Courses?} In: Proceedings of the 2023 Conference on Innovation and Technology in Computer Science Education V. 1. {ACM} (jun 2023). \doi{10.1145/3587102.3588792}

\bibitem{singh2023exploring}
Singh, H., Tayarani-Najaran, M.H., Yaqoob, M.: {Exploring Computer Science students’ Perception of {ChatGPT} in Higher Education: A Descriptive and Correlation Study}. Education Sciences  \textbf{13}(9), ~924 (2023)

\bibitem{wei2022chain}
Wei, J., Wang, X., Schuurmans, D., Bosma, M., Xia, F., Chi, E., Le, Q.V., Zhou, D., et~al.: {Chain-of-Thought Prompting Elicits Reasoning in Large Language Models}. Advances in Neural Information Processing Systems  \textbf{35},  24824--24837 (2022)

\bibitem{xu_systematic_2022}
Xu, F.F., Alon, U., Neubig, G., Hellendoorn, V.J.: {A Systematic Evaluation of Large Language Models of Code}. In: Proceedings of the 6th {ACM} {SIGPLAN} {International} {Symposium} on {Machine} {Programming}. pp. 1--10. ACM, San Diego CA USA (Jun 2022). \doi{10.1145/3520312.3534862}

\bibitem{yilmaz2023augmented}
Yilmaz, R., Yilmaz, F.G.K.: Augmented intelligence in programming learning: {E}xamining student views on the use of {ChatGPT} for programming learning. Computers in Human Behavior: Artificial Humans  \textbf{1}(2),  100005 (2023)

\bibitem{zastudil2023generative}
Zastudil, C., Rogalska, M., Kapp, C., Vaughn, J., MacNeil, S.: {Generative AI in computing education: Perspectives of Students and Instructors}. In: 2023 IEEE Frontiers in Education Conference (FIE). pp.~1--9. IEEE (2023)

\end{thebibliography}
\end{document}